# Effect of molybdenum content on phase transformations during continuous heating of metastable β Ti-Mo alloys

Veronika Kočiščáková[a], Anna Veverková[a,*], Josef Stráský[a], Jana Šmilauerová[a], Kristína Barthа[a], Petr Harcuba[a], Dalibor Preisler[a], Jozef Veselý[a], Miloš Janeček[a]

[a]Department of Physics of Materials, Charles University, Ke Karlovu 5, 121 16, Prague, Czech Republic

## Abstract

Three metastable β titanium alloys with different molybdenum contents, Ti-12Mo, Ti-15Mo and Ti-18Mo (wt.%), were investigated to determine the influence of β-phase stability on phase transformations during continuous heating. Electrical resistance measurement, dilatometry and differential scanning calorimetry were performed during linear heating at 5 °C/min up to 750 °C. The in-situ experiments were complemented by electron microscopy observations of specimens heated to selected temperatures and subsequently water-quenched. All alloys exhibited a sequence involving partial dissolution of athermal ω phase, stabilization and growth of isothermal ω phase, subsequent ω dissolution and precipitation of α phase. Increasing molybdenum content progressively suppressed ω phase evolution and subsequent α precipitation and narrowed the temperature interval associated with ω-related transformations. The excellent agreement between electrical resistance measurements, DSC, dilatometry and electron microscopy enabled a consistent description of phase evolution during heating and demonstrated the benefits of combining complementary experimental techniques for investigation of complex transformation pathways in metastable β titanium alloys.

## 1. Introduction

Titanium and its alloys are widely used in aerospace, biomedical and engineering applications owing to their high specific strength, excellent corrosion resistance and good biocompatibility [1]. Among titanium alloys, metastable β alloys have attracted considerable attention because their microstructure and properties can be tailored over a wide range through thermomechanical processing and controlled phase transformations [2].

The properties of metastable β titanium alloys are strongly influenced by their microstructure, particularly by the type, volume fraction and distribution of secondary phases formed from the parent β phase. During heat treatment, the metastable β phase may transform into several phases, most notably ω and α, which significantly affect mechanical behaviour. Consequently, understanding phase transformation pathways in β titanium alloys is essential for the optimization of their properties and processing routes [1, 2].

Binary Ti-Mo alloys represent a convenient model system for studying phase stability and phase transformations because the stability of the β phase can be controlled simply by varying the molybdenum concentration. Increasing Mo content progressively stabilizes the β phase, suppresses martensitic transformation, reduces the amount of ω precipitates formed after quenching and slows the decomposition of the metastable β matrix during subsequent thermal treatments [3, 4, 5, 6, 7]. Consequently, Ti-12Mo, Ti-15Mo and Ti-18Mo alloys investigated in the present paper provide a suitable compositional series for investigation of the relationship between β phase stability, phase transformation pathways and transformation kinetics.

After solution treatment in the β phase field and quenching, metastable Ti-Mo alloys retain the β phase at room temperature in a thermodynamically metastable state. Regardless of the quenching rate, nm-sized particles of athermal ω ($\omega_{ath}$) phase form in the metastable β matrix [8, 9]. ω phase particles form through a diffusionless collapse of the $(111)_\beta$ lattice planes (bcc), resulting in a hexagonal or trigonal structure depending on the degree of lattice collapse [10]. During subsequent annealing at intermediate temperatures, $\omega_{ath}$ particles transform into larger and more stable isothermal ω ($\omega_{iso}$) precipitates through a diffusion-assisted process accompanied by solute partitioning between the β and ω phases [4, 8, 11]. Atom probe and high-resolution electron microscopy investigations provided new insights into the formation mechanisms of $\omega_{iso}$ and demonstrated the important role of local chemical partitioning during the β→ω transformation [11]. Detailed three-dimensional characterization of ω precipitates in binary Ti-Mo alloys further revealed their nanoscale morphology and chemical composition, highlighting the important role of local chemical fluctuations in ω phase stability [12].

At higher temperatures, ω phase gradually dissolves and α phase precipitates from the β matrix. Numerous studies have shown that ω precipitates can serve as heterogeneous nucleation sites for α phase formation; therefore, the evolution of the ω phase strongly affects the kinetics, morphology and spatial distribution of α precipitates in metastable β titanium alloys. [9, 13].

Significant effort has been devoted to understanding phase transformations in Ti-Mo alloys. Experimental studies have demonstrated that ω precipitation in Ti-Mo alloys is accompanied by both structural and compositional instabilities and is associated with local partitioning of molybdenum between the β matrix and ω precipitates [8]. The transformation sequence involving ω formation, ω dissolution and α precipitation has been investigated in several Ti-Mo alloys using electrical resistance measurements, diffraction techniques and electron microscopy [14, 15, 16, 17, 18, 19]. In parallel, first-principles calculations have provided a thermodynamic explanation for the stabilizing effect of molybdenum on the β phase and for the decreasing tendency toward ω formation with increasing Mo concentration [6, 7]. Phase-field models have subsequently been developed to simulate the nucleation, growth and morphology of ω particles and to investigate the effects of elastic strain and interfacial energy on their evolution [20, 21]. More recently, a crystallographically consistent multiwell phase-field formulation that eliminates non-physical ω–ω interfaces and provides a more rigorous framework for quantitative simulations of β → ω microstructure evolution was proposed in [22].

Among binary Ti-Mo alloys, Ti-15Mo has been studied most extensively. Phase transformations during continuous heating were investigated by in-situ electrical resistance measurements [16, 23], while in-situ neutron and X-ray diffraction experiments provided quantitative information on the evolution of β, ω and α phase fractions during heating [15, 17]. These studies demonstrated that the transformation path is strongly affected by the thermal history and heating conditions. It was shown that dissolution of $\omega_{iso}$ precedes α phase precipitation during heating at a rate of 5 °C/min, while a coexistence of $\omega_{iso}$ phase and α phase was observed on a narrow temperature interval during slower linear heating (1.9 °C/min) [15].

In contrast, considerably less information is available regarding the influence of molybdenum content on transformation behaviour during continuous heating. Although the effect of Mo concentration on β phase stability is generally well established [3, 4, 5, 6, 7], a systematic experimental comparison of phase transformations occurring under identical heating conditions in binary Ti-Mo alloys with different β phase stability is still lacking.

Therefore, the present work investigates phase transformations during linear heating of solution-treated Ti-12Mo, Ti-15Mo and Ti-18Mo alloys. Electrical resistance measurements, dilatometry and differential scanning calorimetry were performed under identical heating conditions in order to compare the sensitivity of individual techniques to ongoing phase transformations. The in-situ experiments were complemented by scanning and transmission electron microscopy observations of specimens heated to selected temperatures and subsequently water quenched. The objective was to determine the influence of molybdenum content on transformation temperatures, phase stability and transformation pathways and to provide a comprehensive comparison of complementary experimental techniques for investigating phase transformations in metastable β titanium alloys.

# 2. Experimental methods

## 2.1 Material

Three binary Ti-Mo alloys containing 12, 15 and 18 wt.% Mo were investigated. The Ti-12Mo and Ti-18Mo alloys were prepared by arc melting under a high-purity helium atmosphere at ÚJP Praha a.s. The resulting ingots weighed approximately 250 g. The Ti-15Mo alloy was supplied by Carpenter Co. (USA) in the form of a forged rod with a diameter of approximately 10 mm.

All alloys were solution treated in the single-phase β region and subsequently water quenched (WQ) to retain the metastable β phase. The forged Ti-15Mo alloy was solution treated at 810 °C for 20 min, which was found to be sufficient for obtaining a fully β microstructure while limiting grain growth. The cast Ti-12Mo and Ti-18Mo alloys already exhibited coarse-grained microstructure and were therefore solution treated at 900 °C for 2 h. The resulting microstructures were free of detectable α phase and consisted of metastable β matrix containing nanoscale $\omega_{ath}$ particles as previously reported for similarly processed Ti-Mo alloys [14, 16, 24].

## 2.2 In-situ electrical resistance measurement

Electrical resistance measurements were employed to monitor microstructural changes during heating. The electrical resistivity of metallic materials is sensitive to structural defects and interfaces that scatter conduction electrons, e.g. solute atoms, vacancies, dislocations, grain boundaries and phase boundaries [25, 26]. Consequently, phase transformations and recovery processes can be detected through changes in the temperature dependence of electrical resistance [25, 27].

The electrical resistance was measured using a four-point method. Samples with dimensions of approximately 15 × 10 × 1 $mm^3$ were cut into an S-shaped meander geometry in order to increase the effective measurement length and improve sensitivity [28]. A custom-built measurement system equipped with a Keithley 2182A nanovoltmeter and a Keithley 2401 sourcemeter was used. Measurements were performed during linear heating at a rate of 5 °C/min in an inert argon atmosphere up to 750 °C.

## 2.3 Dilatometry

Dilatometry was used to investigate dimensional changes associated with phase transformations. Since individual phases differ in lattice parameters, crystal structure and thermal expansion behaviour, ongoing phase transformations can be detected through deviations from the thermal expansion expected for a microstructurally stable material [29].

Samples in the shape of a right prism with dimensions of 3 × 3 × 40 mm3 were measured using a Linseis L75 PT vertical dilatometer. Experiments were carried out in a protective argon atmosphere during linear heating from room temperature to 750 °C at a heating rate of 5 °C/min.

## 2.4 Differential scanning calorimetry

Differential scanning calorimetry (DSC) was employed to detect heat effects associated with phase transformations occurring during heating [30, 31]. Samples weighing approximately 40 mg were prepared from each investigated alloy. Measurements were carried out using a Netzsch DSC 404C Pegasus calorimeter under a protective Ar atmosphere.

The specimens were heated from room temperature to 750 °C at a constant rate of 5 °C/min, identical to that used in the electrical resistance and dilatometric measurements. Throughout this work, the conventional representation is adopted, in which exothermic reactions are displayed as positive heat-flow peaks.

## 2.5 Scanning electron microscopy

Microstructure evolution during heating was investigated using specimens heated to selected temperatures followed by immediate water quenching. Scanning electron microscopy (SEM) observations were performed using a Zeiss Auriga microscope. Backscatter electron (BSE) contrast images are shown in this paper.

Samples for SEM observation were polished using SiC papers up to 2000 grit; subsequently the final polishing step was done using a vibratory polisher with a 0.04 μm colloidal silica suspension.

## 2.6 Transmission electron microscopy

Transmission electron microscopy (TEM) was employed to characterize the evolution of ω and α phases during heating at the nanoscale. Thin foils were prepared from the solution-treated condition and from specimens heated to selected temperatures and subsequently water-quenched. After mechanical grinding, a two-step thinning procedure was applied. First, the specimens were electropolished using a Tenupol-5 (Struers) in a solution of $HClO_4$, butanol and methanol at -20 °C and an applied voltage of 20 V. Final thinning was performed by ion milling in a Precision Ion Polishing System (PIPS, Gatan) operated at 5 kV. To minimize surface damage and improve foil quality, the ion-beam voltage was reduced to 3 kV during the final 30 min of milling. TEM observations were carried out using a JEOL 2200FS transmission electron microscope operated at an accelerating voltage of 200 kV.

# 3. Results

## 3.1 In-situ electrical resistance measurements

The temperature dependence of electrical resistance measured during linear heating at 5 °C/min is shown in Figure 1. To facilitate comparison between the investigated alloys, the resistance is presented relative to its local minimum observed at approximately 572 °C, 560 °C and 545 °C for Ti-12Mo, Ti-15Mo and Ti-18Mo, respectively. Around these temperatures, the microstructure consists predominantly of the β phase, since the ω phase is nearly or completely dissolved while substantial α precipitation has not yet occurred [15]. The electrical resistance curves can therefore be compared in a microstructural state that is similar for all three alloys.

The electrical-resistance behaviour of Ti-15Mo has been thoroughly described in [16] and the interpretations have been supported by in-situ neutron diffraction [15] and in-situ high-energy X-ray diffraction [32] experiments performed under identical heating conditions. These studies provide a useful reference for interpretation of the present results.

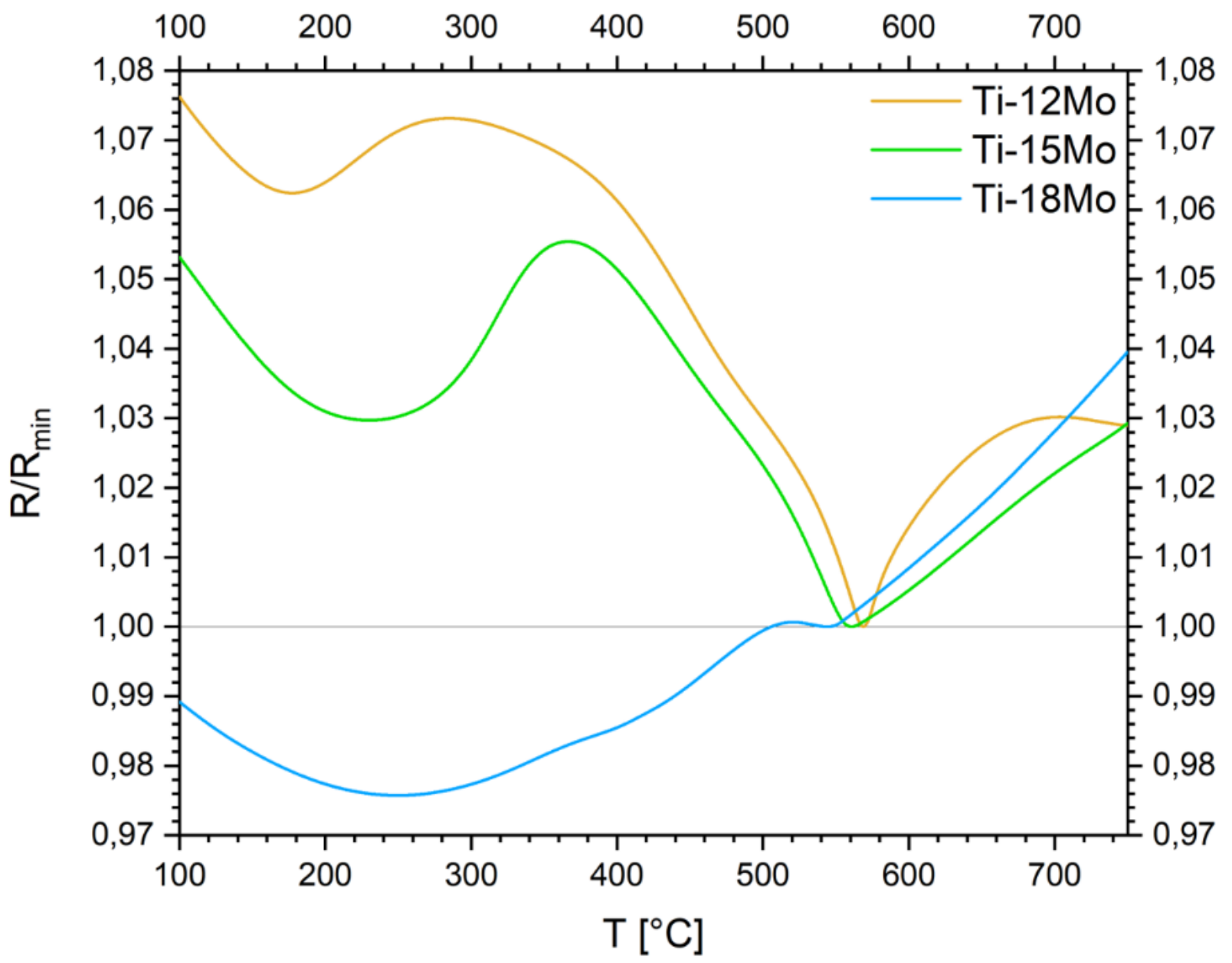

Figure 1: Relative electrical resistance measured during linear heating at 5 °C/min for solution-treated Ti-12Mo, Ti-15Mo and Ti-18Mo alloys. The curves are normalized to the local resistance minima observed at high temperatures. Characteristic features associated with dissolution of athermal ω phase, evolution of isothermal ω phase, dissolution of ω phase and precipitation of α phase are visible. Increasing molybdenum content reduces the magnitude of the transformation-related changes and modifies their temperature range.

All investigated alloys exhibit an initial decrease in electrical resistance immediately after the start of heating. The temperature interval over which this decrease occurs depends on the molybdenum content and extends approximately to 180 °C in Ti-12Mo, 240 °C in Ti-15Mo and 250 °C in Ti-18Mo. The magnitude of the decrease gradually diminishes with increasing molybdenum content and is least pronounced in Ti-18Mo. This initial decrease of electrical resistance has previously been attributed to partial dissolution of athermal ω particles retained after quenching [23, 33].

With increasing temperature, the electrical resistance of all studied alloys starts to increase. This increase reflects not only the normal contribution of phonon scattering but also microstructural changes occurring during ω phase evolution. According to previous diffraction studies of Ti-15Mo, stabilization and growth of $\omega_{iso}$ particles begins approximately between 250 and 300 °C [32]. Formation and growth of these particles increase the number of ω/β interfaces and therefore enhance electron scattering.

However, the increase in electrical resistance is not monotonic. In all alloys, the slope progressively decreases and eventually a local maximum develops. This maximum is located at approximately 280 °C in Ti-12Mo and 360 °C in Ti-15Mo, while in Ti-18Mo it is reduced to only a weak shoulder around 360 °C. The local maxima in Ti-12Mo and Ti-15Mo indicate that the process responsible for decreasing electrical resistance becomes sufficiently intense to overcome the normal increase in resistance caused by phonon scattering. The absence of a true local maximum in Ti-18Mo should not be interpreted as evidence that the corresponding transformation is absent. Rather, it suggests that its contribution to electrical resistance is weaker and never exceeds the opposing contribution arising from phonon scattering.

The precise onset temperature of such a transformation cannot be determined directly from the resistance curve. Since the measured signal reflects the combined contributions of all simultaneously occurring processes, a change in the overall trend becomes observable only when the newly activated process outweighs the others.

The origin of the process associated with decreasing electrical resistance in this intermediate temperature range cannot be unequivocally identified based solely on the resistivity data. Nevertheless, it is most likely associated with the continuing evolution of the ω phase, for example through chemical partitioning between ω and β phases, relaxation of strain fields associated with ω precipitates and/or changes in the character of β/ω interfaces. The substantially weaker manifestation of this feature in Ti-18Mo is consistent with the considerably lower amount of the ω phase in this alloy.

Superimposed on the decrease in electrical resistance is another feature visible as a broad shoulder or a bump centred near 500 °C. The position of this feature is remarkably similar for all three alloys. Above it, the decline in electrical resistance accelerates and culminates in a pronounced minimum. This behaviour is most likely associated with the progressive reduction of the ω phase fraction during its dissolution. Simultaneously, the remaining ω particles become coarser and less numerous [16, 34, 12], reducing the total area of ω/β interfaces and thereby the contribution of the ω phase to electron scattering. Note that ω dissolution, i.e. the process related to the more rapid decline of electrical resistance, outweighs the phonon scattering evolution even in Ti-18Mo – as opposed to the previous process.

The minima located at approximately 572 °C, 560 °C and 545 °C for Ti-12Mo, Ti-15Mo and Ti-18Mo mark the temperature at which the ω phase becomes fully or nearly fully dissolved. The characteristic temperature therefore decreases slightly with increasing molybdenum content. Concurrently, the interval over which ω-related changes can be detected becomes progressively narrower, indicating that increasing molybdenum content modifies not only the amount of ω phase but also its thermal stability.

At higher temperatures, the behaviour of the alloys begins to differ substantially. In Ti-12Mo, a rapid increase in electrical resistance is observed immediately after the minimum. This increase coincides with rapid precipitation and growth of the α phase and reflects both the normal temperature dependence of electrical resistance (i.e. phonon scattering) and additional scattering associated with the formation of a large density of α/β interfaces. In contrast, Ti-15Mo and especially Ti-18Mo exhibit only a weak deviation from the approximately linear increase expected from phonon scattering. The contribution of α precipitation to the electrical resistance signal therefore becomes progressively less pronounced with increasing molybdenum content, consistent with the reduced extent α precipitation in the more strongly β-stabilized alloys.

## 3.2 In-situ differential scanning calorimetry

The DSC curves recorded during linear heating at 5 °C/min are shown in Figure 2. All investigated alloys exhibit several distinct thermal events associated with microstructural evolution during heating. The overall shape of the curves is similar for all compositions; however, the position, intensity and relative prominence of the individual thermal effects depend strongly on the molybdenum content.

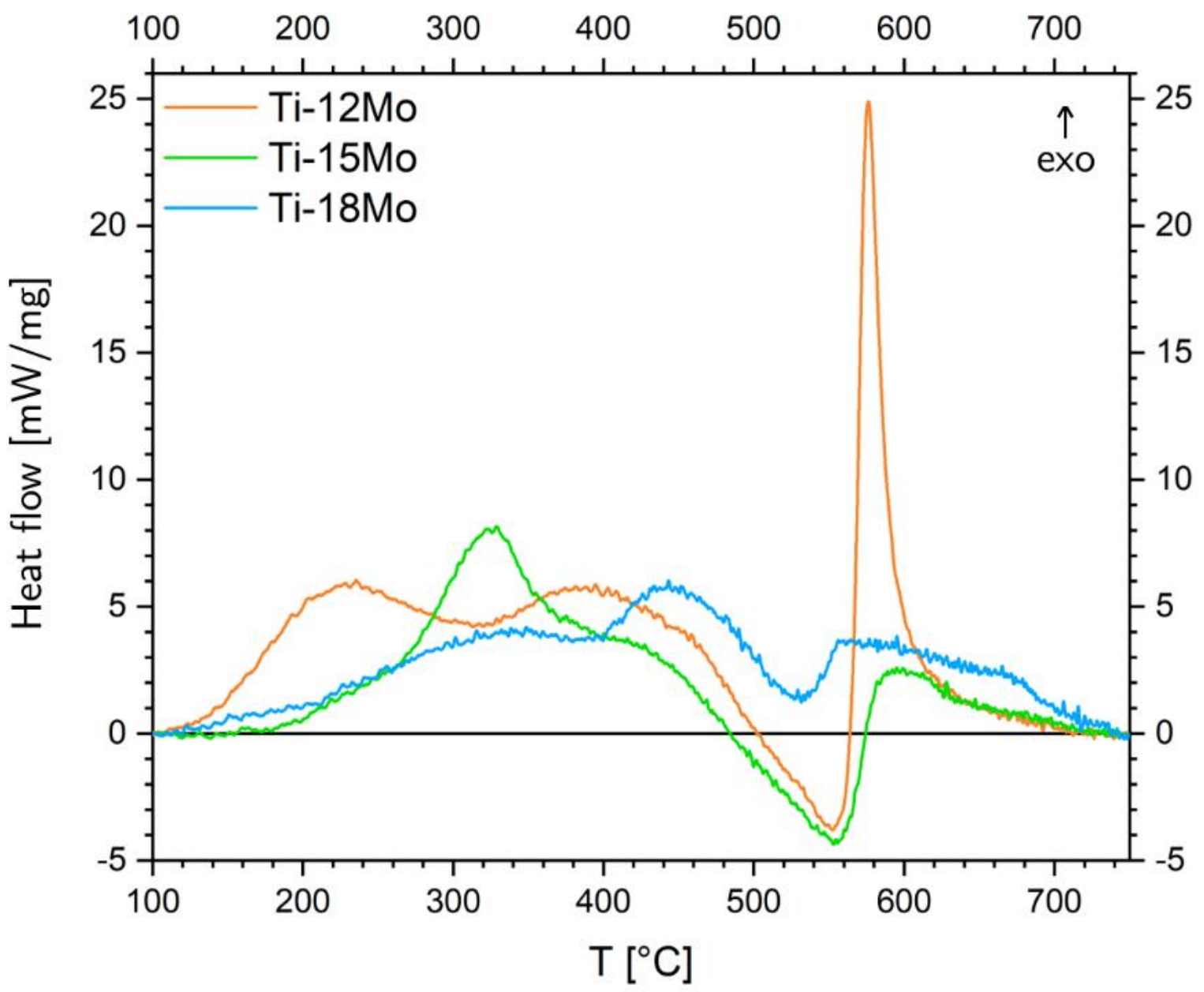

Figure 2: Differential scanning calorimetry curves recorded during linear heating at 5 °C/min. Exothermic heat effects are oriented upwards. The curves were shifted vertically for clarity. Broad exothermic regions associated with ω phase evolution are followed by an endothermic event attributed to ω phase dissolution and a high-temperature exothermic effect associated with α precipitation. The intensity of all thermal effects decreases with increasing molybdenum content.

At intermediate temperatures (approx. 150 °C to 500 °C), broad exothermic regions are observed in all three alloys. Rather than a single peak, these regions appear to consist of two partially overlapping exothermic events. A similar superposition of broad and narrower exothermic features was recently reported during phase transformations in metastable β alloys by Veverková et al. [34].

For Ti-12Mo, the broad exothermic effect extends approximately from 120 °C to 500 °C, while a second superimposed peak can be distinguished between approximately 140 °C and 320 °C. The corresponding features shift towards higher temperatures with increasing molybdenum content. In Ti-15Mo, the broad exothermic event extends approximately from 180 °C to 490 °C, whereas the superimposed peak can be found in the interval between roughly 270 °C and 400 °C. In Ti-18Mo, the broad exothermic peak extends approximately from 160 °C to 520 °C, while an additional exothermic contribution can be distinguished between roughly 400 °C and 500 °C. Notably, the relative positions of the two thermal events differ from those in Ti-12Mo and Ti-15Mo. Whereas in Ti-12Mo and Ti-15Mo the sharper exothermic feature occurs before the maximum of the broad exothermic peak, in Ti-18Mo the broad peak reaches its maximum first and the sharper exothermic event appears at higher temperatures.

The occurrence of two overlapping exothermic processes suggests that the evolution of the ω phase proceeds through more than one microstructural mechanism. Similar behaviour has previously been associated with structural and compositional changes accompanying formation and stabilization of ω precipitates [8, 11, 34]. The broad temperature intervals over which these thermal effects occur indicate that ω evolution is a continuous process rather than a discrete transformation event.

The broad exothermic regions are followed by an endothermic deviation observable in all investigated alloys. The endothermic effect is well pronounced in Ti-12Mo and Ti-15Mo but considerably weaker in Ti-18Mo. The temperature interval of this endothermic signal corresponds closely to the interval in which electrical resistance decreases most rapidly.

At higher temperatures, a final exothermic event is observed. In Ti-12Mo, a sharp peak begins near 560 °C, indicating a relatively rapid transformation. In Ti-15Mo, this peak is substantially weaker and appears only as a broad shoulder. In Ti-18Mo, no distinct exothermic peak can be reliably identified. The intensity of this high-temperature thermal effect therefore decreases systematically with increasing molybdenum content.

## 3.3 In-situ dilatometry

The thermal expansion measured during linear heating is shown in Figure 3. For easier visualization of transformation-related effects, dash-dotted lines representing linear thermal expansion of a microstructurally stable material are included in the figure.

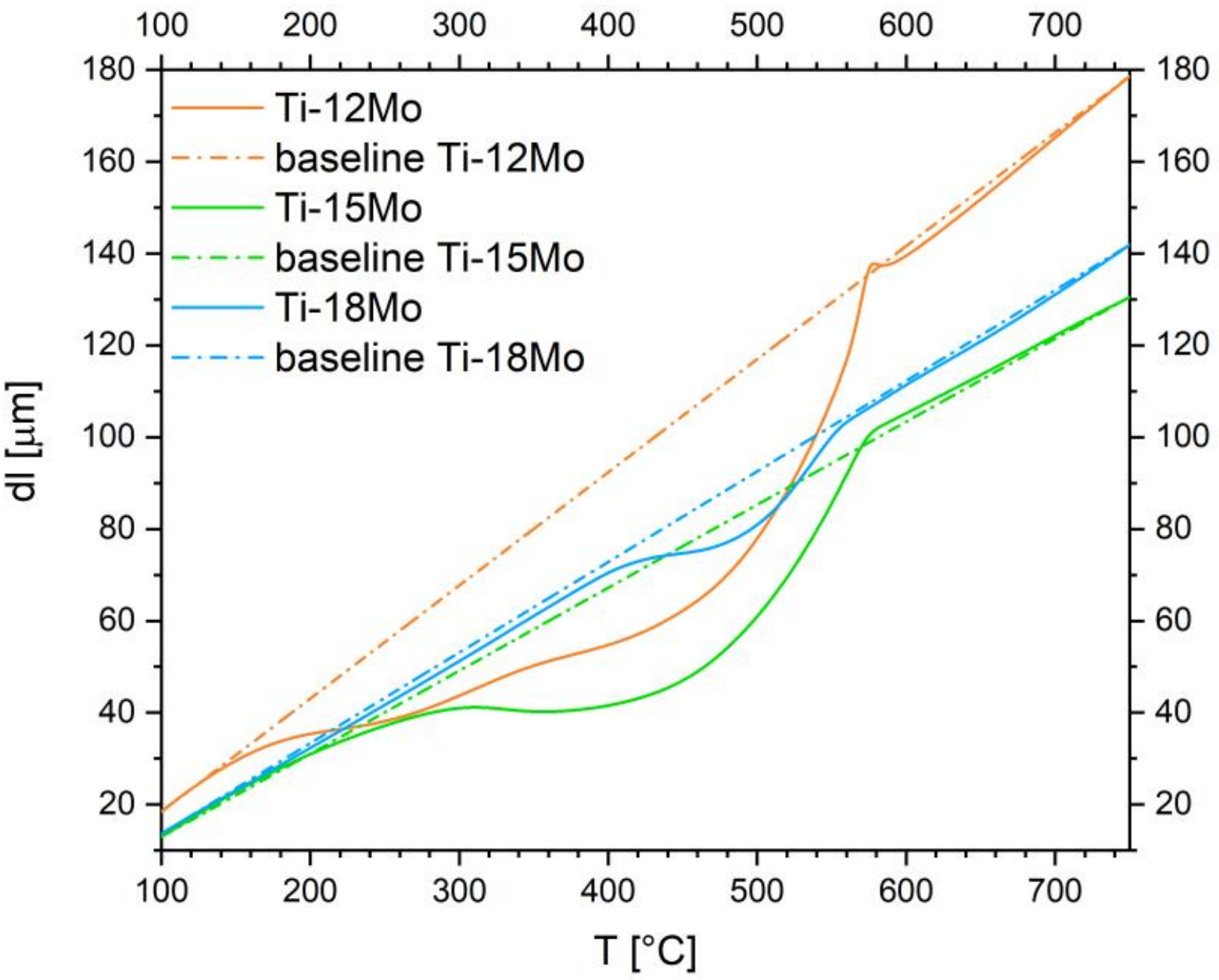

Figure 3: Specimen length change measured by dilatometry during linear heating at 5 °C/min. Dash-dotted lines indicate the approximate thermal expansion expected for a microstructurally stable material. Deviations from the linear trend correspond to transformation-related changes associated primarily with evolution and dissolution of the ω phase.

As reported previously [35, 36], partitioning of molybdenum between β and ω phases results in a reduction of the lattice parameters of both phases, thereby partially counteracting the normal thermal expansion of the material. Consequently, chemical stabilization and growth of ω precipitates counteract the overall elongation associated with thermal expansion. In the dilatometric curves, this behaviour appears as a negative deviation from the linear trend of thermal expansion.

Consistent with this interpretation, all alloys exhibit pronounced deviations from linear thermal expansion over a broad temperature interval, partially counteracting the normal thermal expansion of the material. The onset of this deviation occurs at approximately 130 °C in Ti-12Mo, 180 °C in Ti-15Mo and 170 °C in Ti-18Mo. The curves return to approximately linear trend at around 570 °C for Ti-12Mo and Ti-15Mo and around 560 °C for Ti-18Mo.

A particularly interesting feature of the dilatometric data is the apparent two-stage character of the ω-related transformation interval. This behaviour is most clearly visible in Ti-12Mo. The first deviation from linearity appears at approximately 130 °C, while a second transformation-related feature emerges around 330 °C as an additional reduction in the rate of thermal expansion. A similar two-stage behaviour can also be identified in Ti-15Mo. An initial deviation beginning around 180 °C is followed by a second event appearing slightly below 300 °C. Ti-18Mo exhibits an analogous evolution, although the first deviation starting at around 170 °C is considerably weaker, extending over a wide temperature interval. The second decrease in the slope of the dilatometric curve becomes apparent near 400 °C.

Notably, the temperature intervals associated with the two dilatometric features correspond closely to the positions of the two exothermic events identified in the DSC curves. The agreement between these two fundamentally different experimental techniques indicates that the observed two-stage behaviour represents an intrinsic feature of the transformation process and points towards a complex, multi-stage evolution of the ω phase during heating.

Above approximately 560-570 °C, the dilatometric curves return to a roughly linear character. This behaviour indicates that the processes responsible for the negative deviation from thermal expansion have ceased and is therefore consistent with dissolution of the ω phase.

In contrast to electrical resistance and DSC, dilatometry exhibits only limited sensitivity to α precipitation. A detectable deviation is observed only in Ti-12Mo, where α precipitation is most extensive. For Ti-15Mo and Ti-18Mo, any dilatometric manifestation of α formation remains weak and difficult to distinguish unambiguously from the background thermal expansion.

## 3.4 Scanning electron microscopy

Representative BSE-SEM images of the solution-treated alloys are shown in Figure 4. As expected from their processing history, Ti-12Mo and Ti-18Mo exhibit coarse prior-β grains characteristic of cast material. In contrast, Ti-15Mo possesses a substantially finer grain structure owing to the thermomechanical processing applied by the supplier, as described in the Experimental methods section (Section 2.2).

Microstructures after heating to 600 °C and subsequent water quenching are presented in Figure 5. Significant differences between the investigated alloys are evident. Ti-12Mo exhibits a high density of α lamellae distributed throughout the β matrix. In Ti-15Mo, the amount of α phase is noticeably lower, and the laths are preferentially located near grain boundaries. In Ti-18Mo, only a limited amount of α phase can be identified by SEM.

The observed trend clearly demonstrates that increasing Mo content suppresses decomposition of the metastable β phase and reduces the amount of α phase formed during heating. The trend is fully consistent with the decreasing intensity of the high-temperature DSC peak observed in Figure *2*.

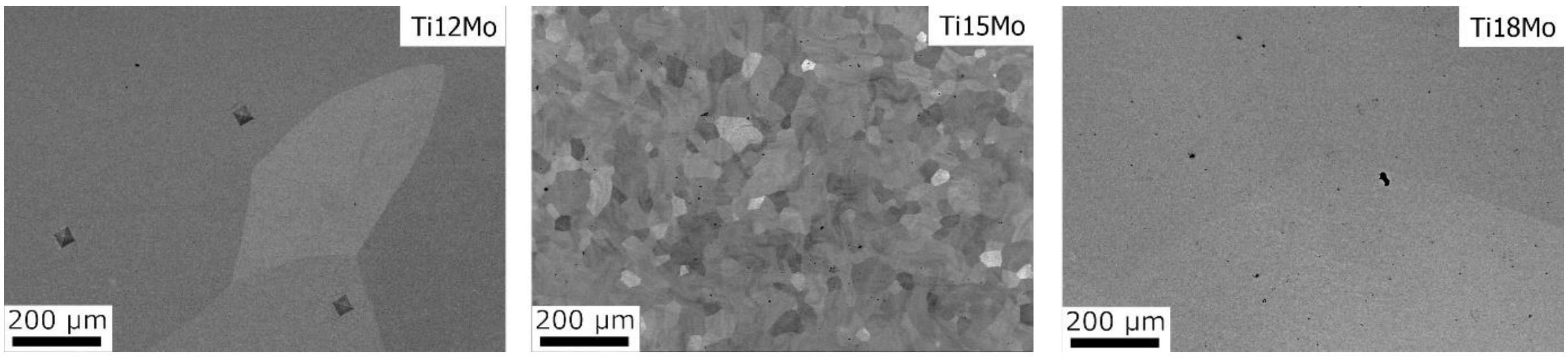

Figure 4: Backscattered-electron SEM images of the solution-treated Ti-12Mo, Ti-15Mo and Ti-18Mo alloys. The cast Ti-12Mo and Ti-18Mo alloys exhibit coarse prior-β grains, whereas Ti-15Mo possesses a considerably finer grain structure due to prior thermomechanical processing. The square-shaped features visible in Ti-12Mo are microhardness indents and are not related to the microstructure.

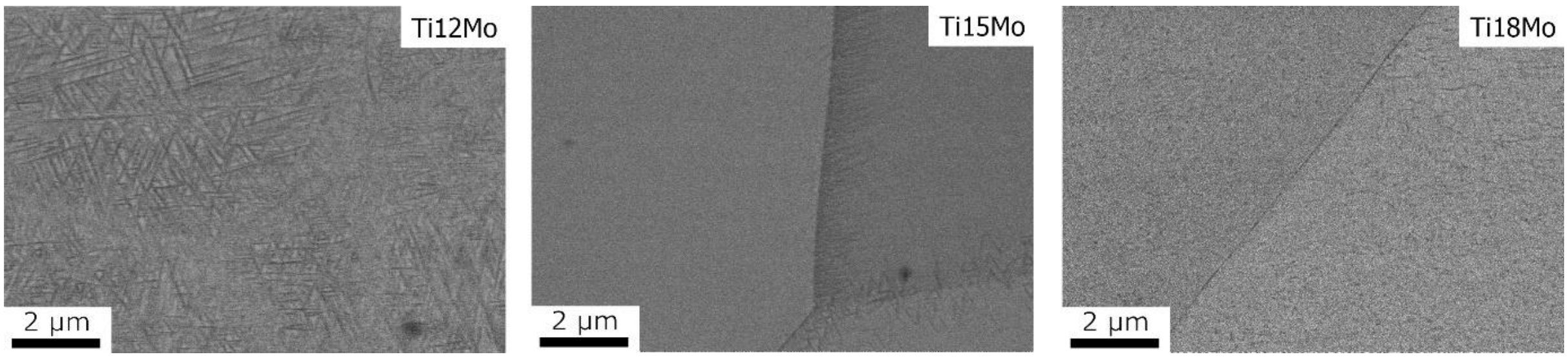

Figure 5: Backscattered-electron SEM images of specimens heated to 600 °C and subsequently water quenched. α precipitates appear as lamellar features within the β matrix. The amount of α phase decreases systematically with increasing molybdenum content.

## 3.5 Transmission electron microscopy

TEM observations of the solution-treated condition are shown in Figure 6. Dark-field (DF) imaging reveals a high number density of extremely fine ω precipitates distributed throughout the β matrix in all investigated alloys. These particles correspond to the athermal ω phase formed during quenching.

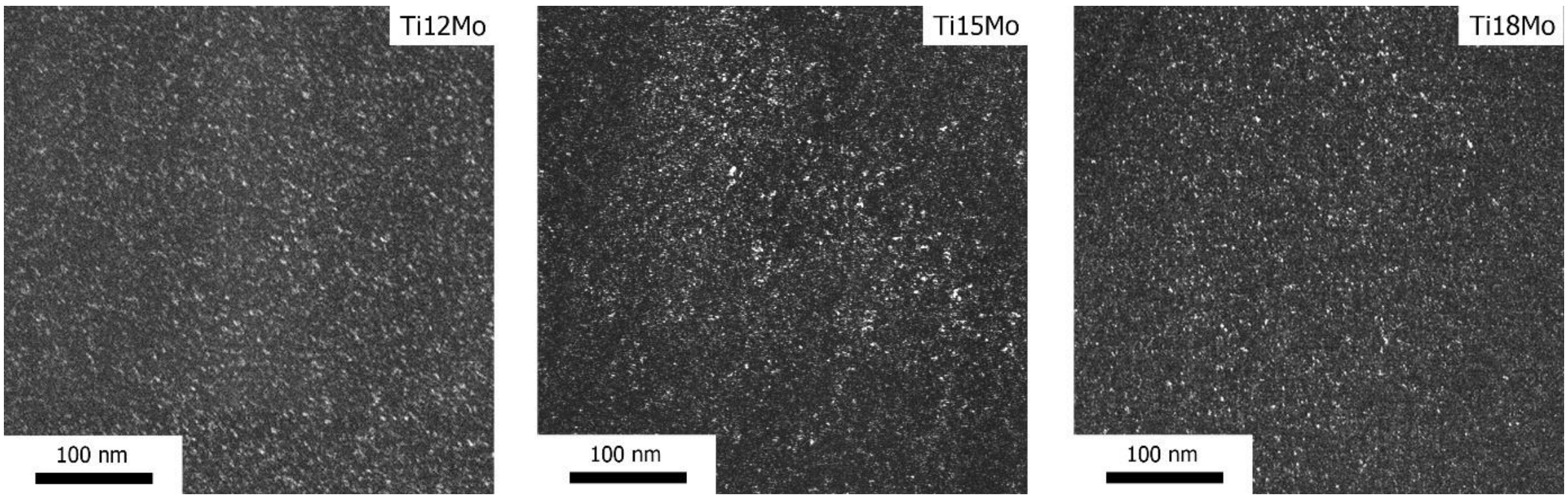

Figure 6: Dark-field TEM images of the solution-treated condition. The images reveal a high number density of nanoscale athermal ω precipitates distributed throughout the β matrix.

After heating to 400 °C and subsequent water quenching (Figure 7), the ω particles become larger than in the solution-treated condition, while their number density decreases. Ti-18Mo exhibits a noticeably lower density of ω precipitates, indicating that formation and growth of ω phase are suppressed by the higher molybdenum content.

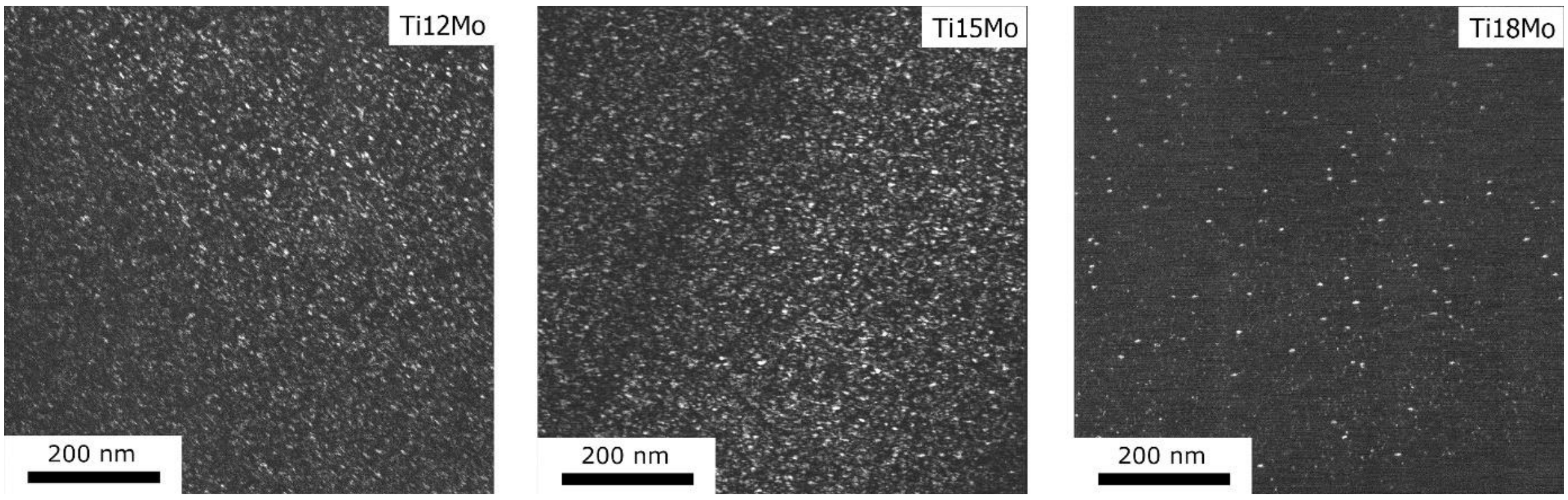

Figure 7: Dark-field TEM images of specimens heated to 400 °C and subsequently water quenched. Compared with the solution-treated condition, ω precipitates are larger and less numerous, indicating significant evolution of the ω phase during heating. The density of ω precipitates decreases with increasing molybdenum content.

Figure 8 shows TEM micrographs of specimens linearly heated to 500 °C and subsequently water quenched. This temperature falls within the initial stage of ω dissolution, as indicated by the onset of the endothermic event in the DSC curves (Figure 2). Compared with the 400 °C condition, the ω particles are considerably coarser in all investigated alloys. Their number density decreased markedly, particularly in Ti-18Mo, where substantially fewer particles are observed than in Ti-12Mo and Ti-15Mo. Interestingly, the average size of the remaining ω precipitates appears comparable in all three alloys despite the pronounced differences in particle density.

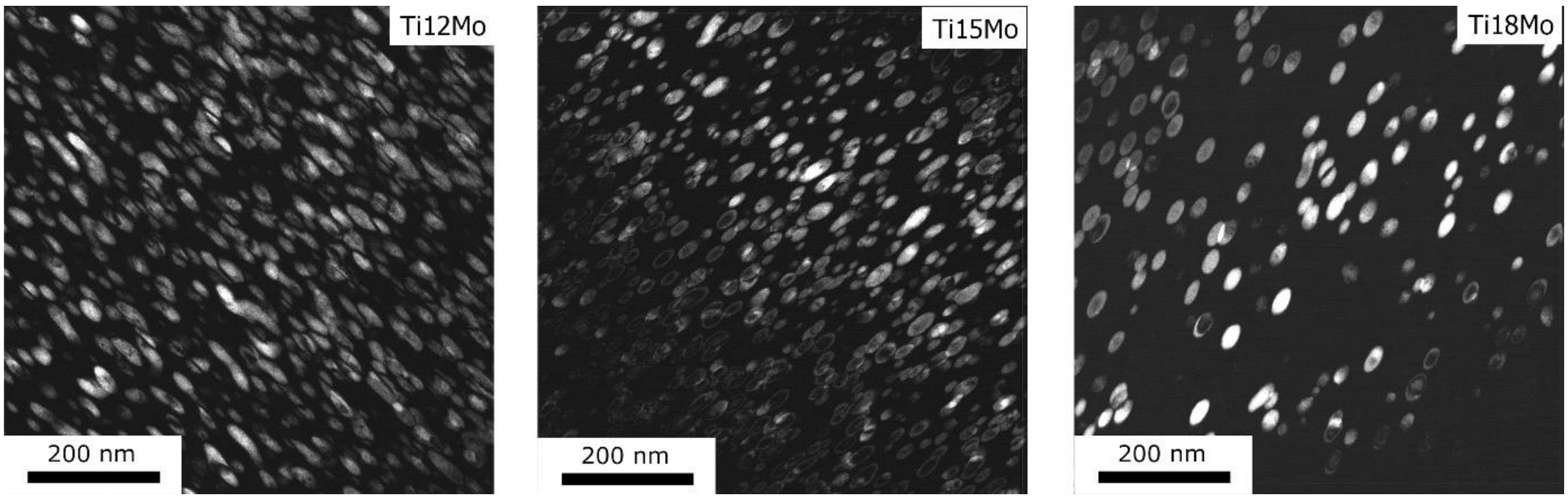

Figure 8 Dark-field TEM images of specimens heated to 500 °C and subsequently water quenched. The ω precipitates are substantially coarser than at 400 °C, while their number density is reduced. This temperature corresponds to the initial stage of ω phase dissolution.

Linear heating to 540 °C (Figure 9) and water quenching results in further coarsening of ω particles and reduction in their number density in Ti-12Mo and Ti-15Mo. In Ti-18Mo, almost no ω particles could be observed, although the electron diffraction pattern showed traces of ω reflections (not shown here). This fact indicates that dissolution of the ω phase is already close to completion in this alloy. This observation agrees remarkably well with the DSC results, where 540 °C falls within the endothermic region associated with dissolution of the ω phase (see the blue curve in Figure *2*).

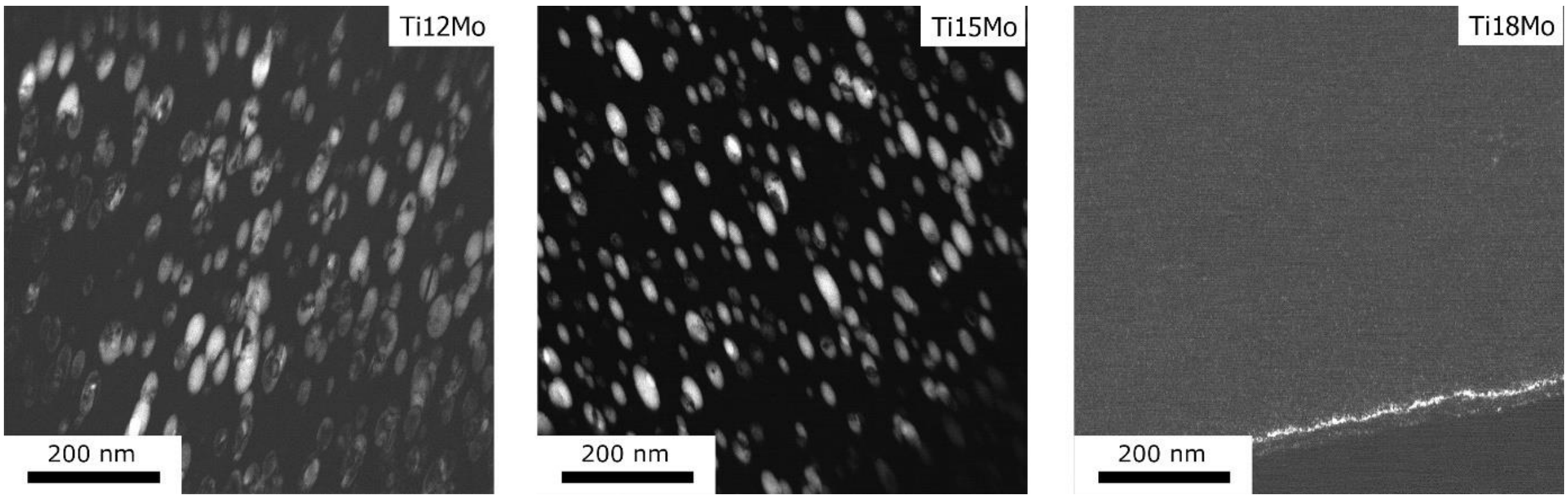

Figure 9: Dark-field TEM images of specimens heated to 540 °C and subsequently water quenched. The number density of ω precipitates is further reduced, particularly in Ti-18Mo, indicating that dissolution of the ω phase is close to completion.

Microstructures after heating to 600 °C and subsequent water quenching are shown in Figure 10. Bright-field (BF) images reveal α precipitates in all investigated alloys. The density of α laths decreases systematically with increasing Mo content, being highest in Ti-12Mo and lowest in Ti-18Mo. While α precipitates in Ti-18Mo are difficult to identify in SEM images (Figure *5*), TEM observations clearly confirm their presence.

The dark-field images in Figure 10 show a fine dispersion of ω particles between the α lamellae. These particles were formed during quenching from 600 °C and therefore correspond to newly formed athermal ω phase rather than remnants of $\omega_{iso}$ present during heating. Similar observations were previously reported for Ti-15Mo quenched during the early stages of α precipitation [16].

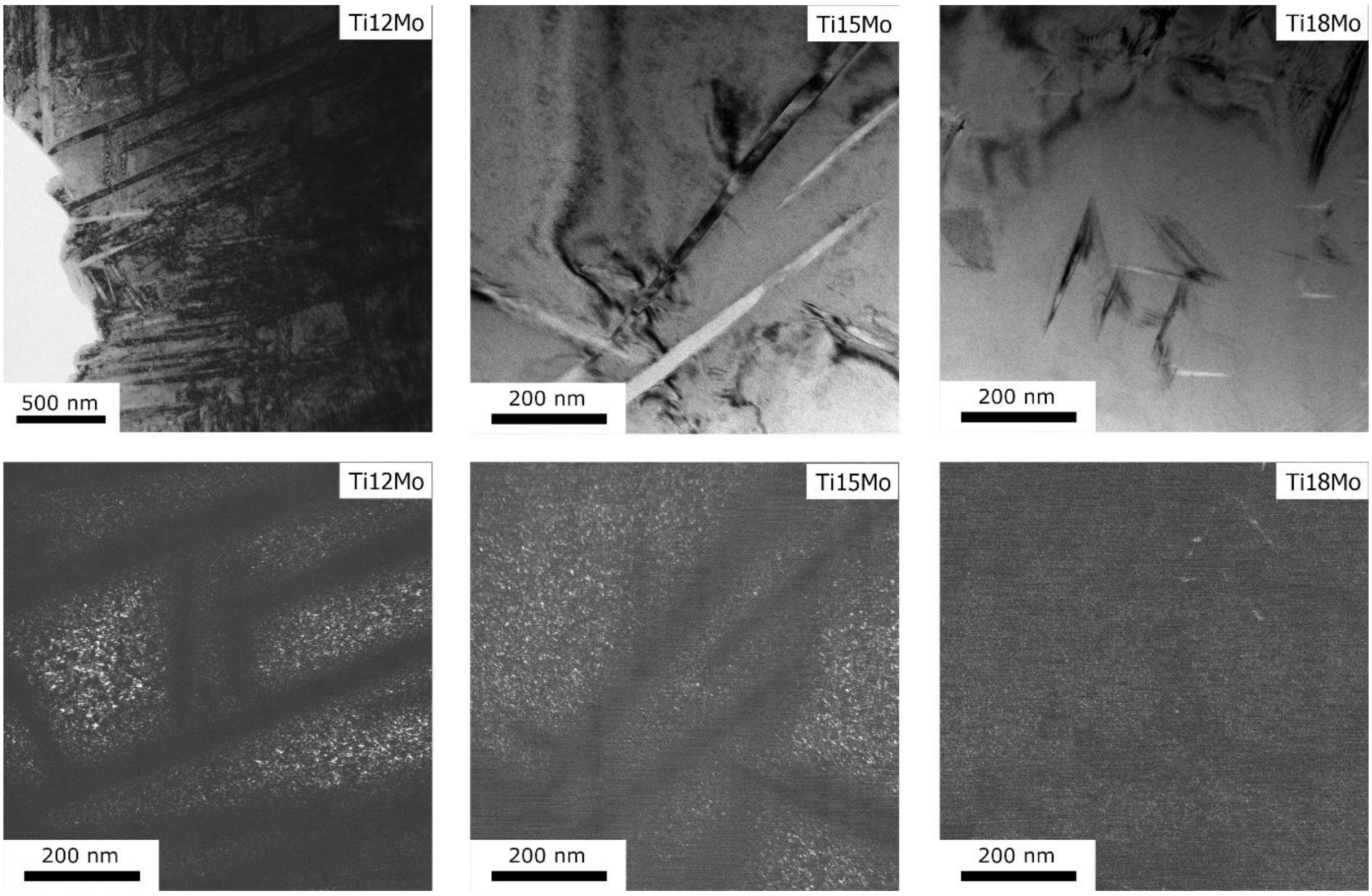

Figure 10: Bright-field (upper row) and dark-field (lower row) TEM images of specimens heated to 600 °C and subsequently water quenched. α precipitates are present in all alloys, while their density decreases markedly with increasing molybdenum content. Fine ω particles visible in the dark-field images correspond to athermal ω phase formed during quenching from 600 °C.

# 4. Discussion

## 4.1 Transformation sequence during heating

The combination of electrical resistance measurements, DSC, dilatometry and electron microscopy provides a detailed picture of phase transformations occurring in the material during heating. Although the individual methods probe different physical quantities, they provide highly complementary information and collectively enable the identification and interpretation of transformation events with considerably greater confidence than any single method alone.

The first transformation-related feature observed during heating is the decrease in electrical resistance extending approximately to 180 °C in Ti-12Mo, 240 °C in Ti-15Mo and 250 °C in Ti-18Mo (see Figure *1*). Similar behaviour was previously observed in Ti-15Mo and attributed to partial dissolution of athermal ω precipitates retained after quenching [17, 33]. The TEM observations presented in Figure 6 confirm the presence of a high number density of nanoscale ω precipitates in the solution-treated state and therefore support the same interpretation. It has been observed that the $\omega_{ath}$ dissolution is a completely reversible process and causes reversed temperature dependence of electrical resistance even far below the room temperature [23, 37]. Despite dissolution of $\omega_{ath}$ having such a profound effect on electrical resistance evolution, DSC and dilatometry unable to detect it. This is not surprising, since electrical resistance is directly tied to changes in electron scattering. Athermal ω precipitates can be regarded as local distortions of the β lattice caused by collapse of selected $\{111\}_\beta$ planes [7, 38]. These distortions act as efficient scattering centres for conduction electrons. Their partial removal during heating increases the electron mean free path and reduces electrical resistance [17]. In all investigated alloys, this effect is sufficiently strong to outweigh the simultaneous increase in resistivity caused by phonon scattering.

Between approximately 150 °C and 500 °C, depending on alloy composition, all investigated techniques indicate continued evolution of the ω phase. Electrical resistance begins to increase (Figure *1*), broad exothermic signals appear in DSC (Figure *2*) and substantial deviations from linear thermal expansion develop in the dilatometric curves (Figure 3). The TEM observations after heating to 400 °C (Figure 7) demonstrate that the ω particles are evidently larger than in the solution-treated condition, confirming extensive microstructural evolution within this temperature interval.

A particularly interesting observation is the occurrence of two partially overlapping transformation events in the temperature interval of approximately 150-500 °C. In DSC, ω evolution is characterized by a broad exothermic signal and a slightly sharper superimposed exothermic contribution (Figure *2*). A remarkably similar two-stage behaviour is independently revealed by dilatometry. The agreement between these techniques is especially noteworthy because they respond to fundamentally different phenomena. DSC reflects heat effects associated with ongoing transformations, whereas dilatometry is sensitive primarily to changes in lattice parameters and thermal expansion behaviour. The fact that both techniques reveal transformation events within the same temperature intervals strongly suggests that the observed features correspond to distinct stages of ω evolution.

The existence of multiple transformation stages is consistent with previous studies of ω formation in metastable β titanium alloys. Devaraj et al. [8] demonstrated that ω formation is accompanied by compositional partitioning between β and ω phases, while Li et al. [11] showed that structural reconstruction and chemical redistribution need not occur simultaneously during β→ω transformation. The broad, two-stage transformation intervals observed in the present study therefore likely reflect coupled structural and compositional processes occurring during ω development rather than a single transformation event.

The electrical resistance data provide additional insight into these processes. Interestingly, the process with the earlier onset around 150 °C with a broad DSC signature has no visible manifestation in the electrical resistance curve. In this temperature interval, only the end of the initial resistance decrease (end of reversible $\omega_{ath}$ dissolution) is observed, followed by an increase of electrical resistance. However, this resistance increase may be largely attributed to normal phonon scattering. On the other hand, the sharper exothermic event in DSC (centred around 220 °C, 320 °C and 450 °C for Ti-12Mo, Ti-15Mo and Ti-18Mo, respectively, see Figure *2*) has its counterparts both in dilatometric and resistometric curves. The sharper DSC events coincide with the first deviation from linear thermal expansion in Ti-12Mo curve, and with the second steps in the dilatometric curves of Ti-15Mo and Ti-18Mo, see Figure 3. Furthermore, the superimposed exothermic feature in DSC coincides approximately with the point at which the rate of the first resistance increase starts to slow down, cf. Figure *1* and Figure *2*. Therefore, this second event must be a process involving an increase of the mean free path of conduction electrons, partially counteracting the phonon scattering. In Ti-12Mo and Ti-15Mo, this process eventually outweighs phonon scattering, while in Ti-18Mo it results only in a broad shoulder just below 400 °C. The exact origin of this process cannot be determined from the present data alone. However, electrical resistance appears to be particularly sensitive to coherent interfaces. A plausible explanation for the observed decrease in electrical resistance is therefore continued evolution of the β/ω interfaces during ω growth. As compositional partitioning progresses and differences between the matrix and precipitates increase, coherency strains may become partially relieved through the formation of semicoherent interfaces [39]. Such a process would partially restore lattice periodicity in the surrounding matrix, increase the mean free path of conduction electrons and consequently reduce electrical resistance. Although direct evidence for this mechanism is not available in the present work, it provides a plausible explanation for the observed decrease in electrical resistance while ω precipitates continue to grow.

The next transformation stage is associated with progressive dissolution of ω phase. This process is most clearly indicated by the pronounced endothermic signal observed in DSC above approximately 500 °C (see Figure *2*). Simultaneously, the electrical resistance exhibits the steepest decline (Figure *1*), and the dilatometric curves return towards the linear thermal expansion behaviour expected for a microstructurally stable material (Figure 3). TEM observations provide direct evidence supporting this interpretation. Compared with the 400 °C condition (Figure 7), specimens heated to 500 °C contain substantially coarser ω precipitates but lower particle densities (Figure 8). After heating to 540 °C, the number density decreases further and ω precipitates become scarce, especially in Ti-18Mo (see Figure 9).

The rapid decrease in electrical resistance during this stage is readily explained in terms of electron scattering. Dissolution of ω particles decreases their volume fraction and simultaneously reduces the total area of β/ω interfaces. As a result, electrical resistance decreases. Unlike the preceding resistance-decreasing process, dissolution of ω phase produces a sufficiently strong effect to outweigh phonon scattering even in Ti-18Mo, resulting in a net decrease in electrical resistance despite increasing temperature.

The final transformation stage corresponds to precipitation of α phase. This process is associated with the high-temperature exothermic peak observed in DSC and is confirmed directly by SEM and TEM observations of specimens heated to 600 °C (Figure *5* and Figure 10). The sequence of ω evolution, ω dissolution and subsequent α precipitation agrees well with previous studies of metastable β titanium alloys [2, 9, 11].

## 4.2 Influence of Mo content

One of the most significant findings of the present study is the systematic influence of molybdenum content on both the extent and character of phase transformations occurring during heating.

All employed techniques indicate that increasing Mo concentration progressively suppresses transformation activity within the metastable β matrix. This trend is already apparent in the electrical resistance curves, where all transformation-related features become noticeably weaker from Ti-12Mo to Ti-18Mo (Figure *1*). The same behaviour is observed in DSC, where both exothermic and endothermic thermal effects decrease substantially in intensity (Figure *2*). Likewise, deviations from linear thermal expansion become progressively less pronounced in the dilatometric measurements (Figure 3).

These observations are fully consistent with the established role of molybdenum as a strong β stabilizer [6, 40, 41]. Increasing Mo concentration lowers the thermodynamic driving force for decomposition of the β phase and therefore suppresses both ω formation and subsequent α precipitation. This trend is directly reflected in the microstructures observed throughout the heating sequence. The density of ω precipitates decreases progressively from Ti-12Mo to Ti-18Mo (see e.g. Figure 8), while the amount of α phase present after heating to 600 °C is dramatically reduced (see Figure *5* and Figure 10). The effect is particularly striking in Ti-18Mo, where only a limited amount of α phase is detected despite undergoing the same thermal treatment.

The present results further suggest that increasing molybdenum content modifies the temperature interval over which ω-related transformations occur. In Ti-18Mo, the transformation-related signals are generally weaker and several features become difficult to distinguish. Interestingly, the characteristic temperatures associated with the early stages of ω evolution shift towards higher temperatures with increasing molybdenum content, whereas the temperature interval corresponding to complete ω dissolution shifts slightly in the opposite direction. This observation indicates that molybdenum affects not only the amount of ω phase formed during heating but also the kinetics of its formation and dissolution. Nevertheless, no evidence was found for a fundamentally different transformation pathway. Instead, all investigated alloys appear to undergo the same sequence of transformations, while increasing molybdenum content progressively reduces the extent of the individual transformation processes.

## 4.3 Comparison of in-situ characterization techniques

The present study also highlights the complementary nature of electrical resistance measurements, DSC and dilatometry for investigation of phase transformations in metastable β titanium alloys. The strengths and limitations of the individual techniques become particularly evident when the complete transformation sequence is considered.

Electrical resistance proved to be the most sensitive technique with respect to subtle microstructural changes. In contrast to DSC and dilatometry, which record thermal and dimensional effects, respectively, electrical resistance responds directly to changes in electron scattering. Consequently, even processes associated with relatively small amounts of transformed material can produce measurable changes in the electrical resistance curves. The initial dissolution of athermal ω phase is a particularly good example. While this process is not visible in DSC or dilatometry, it gives rise to a characteristic decrease in electrical resistance.

DSC provides information that cannot be obtained from electrical resistance measurements. In particular, it can identify phase transformations associated with heat release and heat absorption. The calorimetric data clearly separate the exothermic processes associated with ω evolution and α precipitation from the endothermic process corresponding to ω dissolution.

Dilatometry offers yet another perspective on the transformation behaviour. Owing to its sensitivity to changes in lattice parameters and thermal expansion coefficients, the method responds particularly strongly to processes involving compositional partitioning between β and ω phases. Consequently, the dilatometric curves reveal broad transformation intervals associated with ω evolution and independently confirm the existence of two distinct stages of this process identified by DSC.

Although each individual technique provides valuable information, none of them alone yields a complete description of the transformation sequence. Electrical resistance is highly sensitive but provides only indirect information regarding the nature of the transformations. DSC identifies the energetic character of the ongoing processes but does not directly reveal the underlying microstructural changes. Dilatometry is sensitive to lattice changes but cannot identify the transforming phases. Only when interpreted together and supported by electron microscopy do the individual datasets form a coherent picture of the transformation pathway.

The excellent agreement between electrical resistance measurements, DSC, dilatometry and microscopy observations therefore represents one of the main outcomes of the present work. The combination of these complementary techniques enables phase transformations to be identified and interpreted with substantially greater confidence than would be possible using any of the methods individually, demonstrating the value of a multi-technique approach for investigation of complex transformation pathways in metastable β titanium alloys.

# 5. Conclusions

Phase transformations during linear heating of solution-treated Ti-12Mo, Ti-15Mo and Ti-18Mo alloys were investigated using electrical resistance measurement, dilatometry, DSC and electron microscopy. The following conclusions can be drawn from the present study:

1. All investigated alloys exhibited a transformation sequence involving partial dissolution of $\omega_{ath}$, formation and growth of $\omega_{iso}$, subsequent ω dissolution, and precipitation of α phase.
2. Increasing molybdenum content progressively stabilized the β phase and significantly affected the transformation temperatures and transformation kinetics.
3. The temperature interval over which ω related transformations occurred became narrower with increasing Mo content, and the transformation signatures observed by all in-situ methods became less pronounced.
4. Electron microscopy confirmed a decreasing amount of α phase with increasing molybdenum content, indicating stronger suppression of β phase decomposition in the more highly alloyed materials.
5. Electrical resistance measurements proved highly sensitive to transformations affecting lattice periodicity and the density of scattering centres, particularly those associated with evolution and dissolution of the ω phase. DSC and dilatometry provided complementary information regarding the thermal and dimensional aspects of the transformation sequence, while microscopy enabled direct identification of the evolving phases.

The present study illustrates how information scattered across several experimental techniques can be integrated into a coherent description of phase evolution, providing a powerful framework for investigating complex transformations in metastable alloys.

# Acknowledgements

This work was supported by the Czech Science Foundation (GACR) through the Junior Star project No. 21-18652M. Financial support by the Operational Programme Johannes Amos Comenius of the MEYS of the Czech Republic, within the frame of the project Ferroic Multifunctionalities (FerrMion) [project No. CZ.02.01.01/00/22_008/0004591], co-funded by the European Union is also gratefully acknowledged. V.K. acknowledges the financial support of Grant Agency of the Charles University, project no. 393121.

# Data availability

The data that support the findings of this study are openly available in the Zenodo repository at https://doi.org/10.5281/zenodo.23367020 under the CC-BY 4.0 license.

# Declaration of generative AI and AI-assisted technologies in the manuscript preparation process.

During the preparation of this work, the authors used Microsoft Copilot (accessed October 2026) to improve the readability, grammar and sentence structure of the manuscript. After using this tool, the authors reviewed and edited the content as needed and take full responsibility for the content of the publication.